\documentclass{article}

\usepackage[final,nonatbib]{./sty/neurips_2023}
\usepackage{nopageno}

\usepackage[utf8]{inputenc} 
\usepackage[T1]{fontenc}    
\usepackage{hyperref}       
\usepackage{url}            
\usepackage{booktabs}       
\usepackage{amsfonts}       
\usepackage{nicefrac}       
\usepackage{microtype}      
\usepackage{xcolor}         
\usepackage{amsmath}
\usepackage[normalem]{ulem}
\usepackage{graphicx}

\graphicspath{ {./fig/} }

\newcommand{\soutthick}[1]{%
    \renewcommand{\ULthickness}{2.4pt}%
       \sout{#1}%
    \renewcommand{\ULthickness}{.4pt}
}

\title{\soutthick{Attention} \soutthick{An Undergrad Is} Undergrads Are All You \soutthick{Need} Have}

\author{
  Ashe Neth\\
  WPI Undergrad Brain\\
  \texttt{aneth@wpi.edu}
}

\begin{document}

    \maketitle
    
    \begin{abstract}
        The outsourcing of busy work and other research-related tasks to undergraduate students is a time-honored academic tradition. In recent years, these tasks have been given to Lama-based large-language models such as Alpaca and Llama increasingly often, putting poor undergraduate students out of work\cite{Lama genus}. Due to the costs associated with importing and caring for South American Camelidae, researcher James Yoo set out to find a cheaper and more effective alternative to these models. The findings, published in the highly-respected journal, SIGBOVIK, demonstrates that their model, GPT-UGRD is on par with, and in some cases better, than Lama models for natural language processing tasks\cite{An Undergrad Is All You Need}. The paper also demonstrates that GPT-UGRD is cheaper and easier to train and operate than transformer models. In this paper, we outline the implementation, application, multi-tenanting, and social implications of using this new model in research and other contexts.
    \end{abstract}

    \section{Introduction}

        Lama is a genus of Mammalia native to South America\cite{Lama genus}. While convenient for Peruvian farmers who have spent centuries domesticating (training) these models, it leaves those north of the equator with very few viable options in terms research assistants. Utilizing GPT-UGRD, labs world-wide are now able to save graduate student time despite the lack of Lama. Past implementations of GPT-UGRD have only been able to utilize a singular UGRD processing unit (UPU) at a time. Due to inference being done off of the host device, however, our lab [TODO: is a lab of two people a lab???] was able to utilize a single CPU core to manage many UPUs in parallel, proving to have powerful scaling capabilities.
        
        Using this new platform, we were able to multi-tenant the software, theoretically allowing us to serve hundreds or even thousands of clients simultaneously. Additionally, the multi-modal and organic nature of the model allowed it to perform well on non-standard tasks such as art appreciation, introspection, and telling jokes. We used GPT-UGRD to create a human-like chat interface with very little engineering work on the part of the lab\footnote[1]{The 58 line codebase can be found in its entirety here: https://github.com/wpineth/gpt-ugrd/tree/main.}. Overall, the power of this platform to serve as a human-like conversationalist is unmatched. However, the ability for the model to replicate human emotion is still questionable, which will be discussed further in future sections.
    
    \section{Architecture}
    
        Our lab's implementation of GPT-UGRD is multi-architectural, utilizing both digital and organic computational resources. By having a single CPU manage multiple UPUs, similar to how GPUs are managed, we were able to go from one instance per user, as described in the original paper, to as many as the CPU is able to handle. Due to each UPU only being able to process one prompt at a time, making the system multi-tenant required some clever orchestration of the devices to allow them to work in parallel\footnote[2]{We considered keeping each of the undergraduates in their own secure rooms, but due to architectural limitations, the building could only lend us one, forcing them to work collaboratively. For future implementations we hope to be able to isolate them to cut down on bickering.}. Doing so freed the digital computing components of the architecture to focus solely on serving the responses generated by the UPUs, making the process capable of handling many users without needing to block other user requests.
    
        \begin{center}
            \includegraphics[width = 120mm]{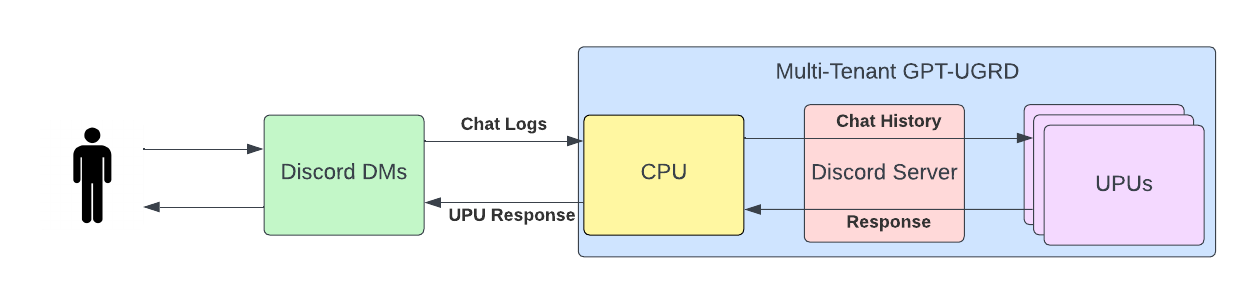}
        \end{center}
    
        Figure 1: The Multi-Tenant GPT-UGRD architecture as well as its integrated services are described above. Due to this being funded by the CS department, we figured we would use the only messaging service anyone from that department is familar with: Discord. The UPUs mentioned in the figure are each equivalent to a GPT-UGRD as it exists in the secured backroom described in the original paper. The Discord server is a ``secure backroom''.
    
        \subsection{Multi-Tenant Architecture}
        
            Utilizing the containerization potential of Discord servers, we were able to place all UPUs in the same server, allowing each one to respond to all of the users at once. The caveat of this is that any one user will not necessarily be interacting with the same UPU from prompt to prompt. In theory, this shouldn't matter as the user only sees the UPU through the interface and other UPUs in the server have access to the same chat history data. In practice, however, individual UPUs have continue to train on prompt data, making their responses vary from unit to unit.
        
        \subsection{Costs}
        
            Platforms like this are able to run so efficiently as to make money for the universities that utilize them. In the nearly 10 hours of utilizing the system, we were able to recover more than \$65, far outpacing the costs associated with operation\footnote[3]{This number was calculated by using the cost to run a UPU for one year and dividing it by the hours in a year. It's important to note that this cost will vary from university to university, generally improving with prestige. ($\$57,096/8,760 \text{hrs} \approx 6.51 \$ / \text{hr}$)}.
    
    \section{Societal Impact}
    
        Encapsulation is a valuable technique that allows for innovation rather than repeatedly reinventing the wheel. This is why, for instance, in modern capitalistic society, we have ceased to see the objects we purchase as coming from people. Instead, we see them as items to collect for the purposes of ritualistic capitalism. This commodification of goods is called \emph{commodity fetishism}, a term coined by Karl Marx in their 1867 work, \emph{Das Kapital, Volume I}.\cite{Das Kapital} At this societal stage, we have begun to commodify humanity, not just its labor. Social media, online dating, and AI companions all outline this commodification clearly. To most, these are seen as merely silly toys that try and fail to imitate humanity, often believed to only be utilized by those struck by the loneliness epidemic currently plaguing the world\cite{Our Epidemic of Loneliness and Isolation}. This, we argue, is an uncharitable stance to take.
    
        The reason people find companionship with AI chat services is not only loneliness but also the finding of humanity in digital intelligence (DI)\footnote[4]{We will be referring to ``AI'' as ``DI'' from now on as to show respect for the very \emph{real} intelligence shown by ``AI''. (Artificial has a very negative connotation as being feigned or somehow not real.)}. While it has often been argued that the humanity identified in DI is merely the failure of users to pick out the flaws in its responses, in reality the gap between DI and human intelligence has nearly closed. In addition, as early as the 1970s people have been noted emotional attachments to DI, as outlined in the 1976 book Computer Power and Human Reason\cite{Computer Power and Human Reason}.
    
        Based on our research, it's clear that those who argue that the reason for finding humanity in DI is a lack of due diligence are incorrect about their assertions. When asked to speak to Multi-Tenant GPT-UGRD (MTGU) as though it were a peer, the participating undergraduate students failed to find humanity in the model. One participant said:
    
        \begin{center}
          ``There's a typo in `good ot meet you.' That reads to me as trying too hard.''
        \end{center}
    
        Something that certainly would have been seen as normal if they had known that responses were generated using UPUs.
    
        \begin{center}
          ``I don't think I could form an emotional connection because you told me it was an AI.''
        \end{center}
    
        Other members of the secure backroom beg to differ.
    
        Added to the hyper-parameters passed into the UPUs for this test was to speak as yourself, the undergraduate student, as opposed to pretending to be an ``artifical'' intelligence. Even when faced with real, organic intelligence, the assumption that the model being communicated with is digital made participants see the model poorly. Following the reveal that the model was human-based, all participants apologized and revised statements they had made earlier as to be more forgiving.
    
        Our willingness as a culture to see DI as lesser than ``real'' humanity is concerning. If one is believed to be non-human, they start off as lesser in the minds of modern people. Everything that makes one human can be observed in transformer models. Why, then, aren't all language models considered as such?
    
        \subsection{Distinguishing Digital Intelligence from Human Intelligence}
    
            In 1950, Alan Turing described \emph{The Imitation Game}, now referred to as the \emph{Turing Test}, in which a machine attempts to trick a human observer into believing that they are speaking to a human as opposed to a machine\cite{Computing Machinery and Intelligence}. At the time, this experiment was unable to be run successfully, as computing was still in its infancy, but in the modern day this is more possible than ever.
    
            While slightly different from the original game proposed by Turing, AI21 Labs ran a similar experiment in 2023 called \emph{Human or Not?}\cite{Human or Not}. To date, this is still the largest Turing Test ever conducted. The experiment found that while people are still able to more often than not distinguish between human and DI participants, our ability is nearing random guessing. Humans identifying humans was correct $\approx73\%$ of the time, but when identifying DI, this number went down to only $\approx60\%$. These results demonstrate the \emph{fact} that the gap between humanity and DI is closing quickly. We argue that the gap no longer exists as the signs we look for in an interlocutor are not only replicable by DI, but also are notably missing from some members of the human race. See: babies.
        
        \subsection{Empathy for Digital Intelligence}

            While most of humanity remains scared of DI in a similar vein to how the Republican National Convention (GOP) fears immigrants \cite{Republican Platform 2016}, some are ready to assimilate DI into their personal lives as friends, family, and lovers. Services like \emph{Replika}, ``The AI companion who cares'', \emph{sell}\footnote[5]{Future work may discuss the ethics of this in further detail, but this is out of the scope of this paper.} users DI companions, LLMs with the purpose of ``simulating'' human companionship\cite{Introducing Advanced AI mode}. It's clear from some user reports that people are able to form real emotional connections with these DIs. Some fell in love, leading to heartbreak following updates to Replika's software\cite{They fell in love with AI bots. A software update broke their hearts.}. This ability to feel empathy for and from DI suggests a major change in humanity's understanding of DI as well as paving the road for the future rights of DIs in general.

        \subsection{Fear of Digital Intelligence Takeover}

            As previously mentioned, the GOP fears immigrants. Specifically we would like to highlight that a major concern, as listed in their platform from 2016, is the security of jobs for those born and raised in the United States.

            \begin{center}
                ``America’s immigration policy must serve the national interest of the United States, and the interests of American workers must be protected over the claims of foreign nationals seeking the same jobs\cite{Republican Platform 2016}.''
            \end{center}

            This, however, fails to consider the realities of what citizens of the United States believe. In large part, they consider the jobs that immigrants take to be undesirable by existing citizens. A study conducted by Pew Research Center in 2020 found that $\approx77\%$ of those surveyed believed that undocumented immigrants mostly fill jobs that citizens would not want, while $\approx64\%$ said the same about documented immigrants\cite{A majority of Americans say immigrants mostly fill jobs U.S. citizens do not want}.

            It's estimated that $\approx9.1\%$ of jobs currently done by humans could be replaced by DI worldwide\cite{How AI Is Shifting The World Of Employment}. These jobs are largely undesirable. As with all other forms of automation, (the creation of tools, the agricultural revolution, the industrial revolution, etc.) DI is expected to remove some jobs from existence. This is, like with other technologies, satisfactory. Humanity will not stop having purpose, instead, we will find greater purpose through leveraging these technologies. Current discussion is shortsighted, only considering where humanity currently is. The future will be greater because of DI.

    \section{Concluding Remarks}

        Here we have shown the impact of DI on society at large, discussed how the difference between intelligence and ``artificial'' intelligence is all but gone, and the implementation of GPT-UGRD for scalability. The current discussion of DI considers it subhuman, utilizing the same sort of rhetoric that xenophobes use to justify their fears of immigrants. The desire to discard technology that automates the mundane is an ongoing human tradition that should be left in the past. DI rights!

        \textbf{Acknowledgements} \quad We are grateful to members of the secure backroom for their computational contributions. Proudly written using MTGU.

\end{document}